# A Unified Approach to Simultaneous Testing of Superiority and Non-inferiority on Multiple Endpoints in Clinical Trials

Wenfeng Chen[1], Naiqing Zhao[2], Guoyou Qin[2], and Jie Chen[3]*

[2]Biostatistics and Statistical Programming
Merck Serono (Beijing) Pharmaceutical R & D Co., Beijing 100022, China

[3]Department of Biostatistics, School of Public Health
Fudan University, Shanghai, 200032, China

[3]Data Science and Pharmacovigilance, ECR Global
Shanghai, China

September 29, 2023

*Correspondence to: Jie Chen, ECR Global, E-Mail: jiechen0713@gmail.com

**Abstract**

Simultaneous tests of superiority and non-inferiority hypotheses on multiple endpoints are often performed in clinical trials to demonstrate that a new treatment is superior over a control on at least one endpoint and non-inferior on the remaining endpoints. Existing methods tackle this problem by testing the superiority and non-inferiority hypotheses separately and control the Type I error rate each at $\alpha$ level. In this paper we propose a unified approach to testing the superiority and non-inferiority hypotheses simultaneously. The proposed approach is based on the UI-IU test and the least favorable configurations of the combined superiority and non-inferiority hypotheses, which leads to the solution of an adjusted significance level $\alpha'$ for marginal tests that controls the overall Type I error rate at pre-defined $\alpha$. Simulations show that the proposed approach maintains a higher power than existing methods in the settings under investigation. Since the adjusted significance level $\alpha'$ is obtained by controlling the Type I error rate at $\alpha$, one can easily construct the exact $(1 - \alpha)\%$ simultaneous confidence intervals for treatment effects on all endpoints. The proposed approach is illustrated with two real examples.

# 1 Introduction

Clinical trials often use multiple endpoints to capture various therapeutical benefits of an investigational product over a (active) control. For instance, randomized trials conducted in patients with Alzheimer's disease usually employ Alzheimer's disease assessment scale-cognitive subpart (ADAS-Cog) and clinical interview-based impression of change (CIBIC-Plus) as measures of treatment effects of a drug (Vellas et al., 2008), and asthmatic studies commonly use forced expiratory volume in one second ($FEV_1$), peak expiratory flow rate (PEFR), symptoms score (SS, in 0-6 scale) and additional medication use (AMU) (e.g., $\beta$-agonist) to indicate disease progression or improvement as a result of medical treatment (Zhang et al., 1997; FDA, 2016). As a generic example, a clinical trial may use both efficacy and safety as bivariate endpoints with one endpoint showing superiority and the other non-inferiority in order to reach a positive conclusion and thereby gain a regulatory approval. A trial is concluded to be positive or successful if the new treatment is shown to be superior over the control on at least one endpoint and non-inferior on the remaining endpoints via a multiple testing procedure that controls the Type I error probability.

Several methods have been proposed for testing superiority and non-inferiority of multiple endpoints in clinical trials. Bloch et al. (2001) introduce a bootstrap-based approach to multiple-endpoint (one-sided) testing that incorporates univariate and multivariate treatment effects. A non-parametric version of Bloch et al. (2001) is developed by Bloch et al. (2007) to demonstrate that the new treatment is non-inferior to the control for all endpoints using marginal comparisons, and is superior to the control for some endpoints using a global comparison. Bloch et al. (2007) point out that their proposed bootstrap method not only handles the complexity of the null hypotheses and their associated test statistics, but also incorporates various dependence in the multivariate distributions of treatment effects (endpoints). Perlman and Wu (2004) replace the classical Hotelling $T^2$ test in Bloch et al. (2001) with the likelihood ratio test that utilizes the appropriate multivariate one-sided test. Tamhane and Logan (2004) develop a method that is based on the union-intersection (UI) principle of Roy (1953) for testing superiority on some endpoints and the intersection-union (IU) principle of Berger (1982) for testing non-inferiority on all endpoints and thereby refer this method as the UI-IU test. They recognize that the critical values for IU tests of non-inferiority may not be further reduced, the constants for UI tests of superiority can be sharpened based on least-favorable configuration of the null hypotheses for superiority. Tamhane and Logan (2004) point out that the UI-IU test is conservative because it requires the Type I error to be controlled separately when testing superiority and non-inferiority of multiple endpoints. Röhmel et al. (2006) argue that it is unacceptable that different choices of non-inferiority margins may impact the conclusion of superiority test and hence propose for a bivariate-endpoint case a three-step procedure in which non-inferiority for both endpoints is tested in the first step, then superiority is tested using a bivariate test (e.g., Holm (1979) or Hochberg (1988)) or Läuter (1996)'s one-sided standard sum (SS) test in the second step, and finally superiority is tested on each endpoint using univariate test in the third step.

As discussed above, all the proposed statistical methods test the superiority and non-inferiority hypotheses separately, each at pre-defined significance level $\alpha$, which does not fully utilize the information on the relationship of test statistics including their dependence structure and superiority and non-inferiority margins. In this paper, we develop a unified approach to simultaneously testing the null hypotheses of superiority and non-inferiority. Specifically, we formulate the problem based on Tamhane and Logan (2004)'s UI-IU test and use the least favorable configurations of the combined null hypotheses of superiority and non-inferiority, which provides the basis for determination of the upper bound of the overall Type I error rate that depends on not only the least favorable configurations of the null hypotheses and the correlation of corresponding test statistics, but also the combined standardized superiority and non-inferiority margins of individual endpoints. Regarding the correlation among endpoints, we can either plug in the estimated partial correlation coefficients, possibly unequal, or use Armitage and Parmar (1986)'s method to approximate the mean partial correlation coefficient which simplifies the computation of critical values for the test statistics.

The rest of the paper is organized as follows. Section 2 describes the problem, preliminaries and notations for testing superiority and non-inferiority hypotheses. In section 3 we present a unified approach to simultaneous testing of superiority and non-inferiority with a focus on the control of overall Type I error rate and the determination of critical values. Simulation results are given in Section 5 to compare our approach to existing methods in terms of Type I error rate control and power. Two examples are presented in Section 6 to illustrate the proposed approach. Finally, some discussions and concluding remarks are given in Section 7.

## 2 Preliminaries and Notations

Consider a two-arm, randomized clinical trial comparing a treatment with a control on $m \geq 2$ endpoints. The objective is to demonstrate superiority of the treatment over the control on at least one endpoint and non-inferiority on the remaining endpoints. Following the notations in Tamhane and Logan (2004), let $\mathbf{X}_{ij} = \{X_{ijk} : k = 1, \ldots, m\}$ be a vector of response variables from the $j$th subject in the $i$th group, $i = 1, 2$ and $j = 1, \ldots, n_i$. Suppose that $\mathbf{X}_{ij}$ are independent and identically distributed from an $m$-variate normal distribution with mean vectors $\boldsymbol{\mu}_i = \{\mu_{i \cdot k} : k = 1, \ldots, m\}$ and a common unknown variance-covariance matrix $\Sigma = \{\rho_{k_1 k_2} \sigma_{k_1} \sigma_{k_2}\}$, $1 \leq k_1 \leq k_2 \leq m$. Let $\boldsymbol{\theta} = \{\theta_k : k = 1, \ldots, m\}$ with $\theta_k = \mu_{1 \cdot k} - \mu_{2 \cdot k}$ being the mean difference in the $k$th endpoint between the two groups. The treatment is considered superior over the control if $\theta_k \geq \epsilon_k$ or non-inferior to the control if $\theta_k \geq -\eta_k$ in the $k$th endpoint, where $\epsilon_k \geq 0$ and $\eta_k \geq 0$ are suitably pre-defined respective superiority and non-inferiority margins for the $k$th endpoint.

Let's define the null hypothesis of superiority $H_{0k}^S : \theta_k \leq \epsilon_k$ and its alternative hypothesis $H_{1k}^S : \theta_k > \epsilon_k$, and the null hypothesis of non-inferiority $H_{0k}^{NI} : \theta_k \leq -\eta_k$ and the corresponding alternative hypothesis $H_{1k}^{NI} : \theta_k > -\eta_k$ for the $k$th endpoint, $k = 1, \ldots, m$. Let $H_0^S : \cap_{k=1}^m H_{0k}^S$ and $H_0^{NI} : \cup_{k=1}^m H_{0k}^{NI}$, then $H^S = \cup_{k=1}^m H_{1k}^S$ and $H^{NI} : \cap_{k=1}^m H_{1k}^{NI}$. The objective of the trial can be formulated as testing the null hypothesis

$$H_0 : H_0^S \cup H_0^{NI}$$

against the alternative hypothesis

$$H_1 : H_1^S \cap H_1^{NI},$$

which is termed as UI-IU test in Tamhane and Logan (2004). Let $\bar{X}_{i \cdot k} = \frac{1}{n_i} \sum_{j=1}^{n_i} X_{ijk}$ be the sample mean of the $k$th endpoint for the $i$th group, $i = 1, 2$, and

$$T_k = \frac{\bar{X}_{1 \cdot k} - \bar{X}_{2 \cdot k} - \theta_k}{\hat{\sigma}_k \sqrt{1/n_1 + 1/n_2}}, \quad (1)$$

with $\hat{\sigma}_k$ being an estimate of $\sigma_k$. Each of the $T_k$'s marginally follows a univariate $t$-distribution with $d = m(n_1+n_2-2)$ degrees of freedom, denoted by $t_d$, and $\mathbf{T} = (T_1, \ldots, T_m)$ follows an $m$-variate $t$-distribution with degrees of freedom $d$ and partial correlation coefficients $\rho_{k_1 k_2}$.

To test $H_{0k}^S : \theta_k \leq \epsilon_k$ against $H^S : \theta_k > \epsilon_k$, one simply constructs the test statistic $T_k^S$ by replacing $\theta_k$ in (1) with $\epsilon_k$. Similarly, one obtains the test statistic $T_k^{NI}$ for testing $H_{0k}^{NI} : \theta_k \leq -\eta_k$ against $H_{1k}^{NI} : \theta_k > -\eta_k$ by replacing $\theta_k$ in (1) with $-\eta_k$. Define

$$\eta_k'(\theta_k) = \frac{\eta_k + \theta_k}{\hat{\sigma}_k \sqrt{1/n_1 + 1/n_2}} \quad \text{and} \quad c_k = \frac{\epsilon_k + \eta_k}{\hat{\sigma}_k \sqrt{1/n_1 + 1/n_2}}$$

for $k = 1, \ldots, m$. Then the test statistics $T_k^{NI}$ can be expressed in terms of $T_k$ and $\eta_k'(\theta_k)$

$$T_k^{NI} = T_k + \eta_k'(\theta_k) \quad (2)$$

and $T_k^S$ in terms of $T_k^{NI}$ and $c_k$

$$T_k^S = T_k^{NI} - c_k, \qquad (3)$$

$k = 1, \ldots, m$. Under the null hypothesis $H_{0k}^{NI}$, i.e., $\theta_k \leq -\eta_k$, one has $\eta'(\theta_k) \leq 0$. Hence the null hypothesis $H_{0k}^{NI}$ is rejected if

$$T_k^{NI} = T_k > t_{d,\alpha'}, \qquad (4)$$

and $H_{0k}^S$ is rejected if

$$T_k > t_{d,\alpha'} + c_k, \qquad (5)$$

where $t_{d,\alpha'}$ is the $100(1 - \alpha')\%$ percentage point of the $t_d$ with $\alpha'$ being appropriately adjusted for multiple tests in order to control the overall Type I error probability at $\alpha$.

Tamhane and Logan (2004) propose to test $H_0^S$ against $H_1^S$ and $H_0^{NI}$ against $H_1^{NI}$ separately, each at level $\alpha$, and the overall Type I error for testing $H_0$ against $H_1$ is controlled at $\alpha$. They develop a method to obtain sharper critical values for the UI-IU test using bootstrap algorithm in a stepwise manner. In the next section, we develop a unified approach for simultaneous tests of superiority and non-inferiority where critical values are obtained using expanded probability expression for the Type I error rate based on least favorable configurations of the combined superiority and non-inferiority null hypotheses.

## 3 A Unified Approach

We first present an improved upper bound of Type I error probability based on which the adjusted significance level $\alpha'$ and its corresponding critical values that control the overall Type I error rate are derived.

## 3.1 Type I error probability

With the rejection regions defined marginally for $H_{0k}^{NI}$ and $H_{0k}^{S}$ as in (2) and (3), the overall Type I error probability $\gamma$ for testing $H_0$ against $H_1$ is hence given by

$$\gamma = P_0\left\{ \cup_{k=1}^{m}(T_k > t_{d,a'} - \eta_k'(\theta_k) + c_k) \cap \cap_{k=1}^{m}(T_k > t_{d,a'} - \eta_k'(\theta_k)) \right\}, \tag{6}$$

where $P_0\{\cdot\}$ denotes the probability under the null hypothesis $H_0$. By the improved Bonferroni inequality of Worsley (1982), an upper bound of $\gamma$ can be written as

$$\begin{aligned}
\gamma \leq &\sum_{k=1}^{m} P_0\left\{ T_k > t_{d,a'} - \eta_k'(\theta_k) + c_k \cap \cap_{i=1}^{m \neq k}(T_i > t_{d,a'} - \eta_i'(\theta_i)) \right\} \\
&- \sum_{k=1}^{m,k_1} P_0\left\{ (T_{k_1} > t_{d,a'} - \eta_{k_1}'(\theta_{k_1}) + c_{k_1}) \cap (T_k > t_{d,a'} - \eta_k'(\theta_k) + c_k) \right. \\
&\left. \cap \cap_{j=1}^{m \neq k_1,k}(T_j > t_{d,a'} - \eta_j'(\theta_j)) \right\},
\end{aligned} \tag{7}$$

in which we use the fact that $T_k > t_{d,a'} - \eta_k'(\theta_k) + c_k$ implies $T_k > t_{d,a'} - \eta_k'(\theta_k)$. Denote the right-hand side of (7) by $\gamma'$. It is not immediately clear how $\gamma'$ behaves in the domain of individual $\theta_k$'s. The following lemma states the relationship of $\gamma'$ with any single $\theta_k$, a necessary condition which helps prove the main result presented in Theorem 1.

**Lemma 1.** *Conditional on $\boldsymbol{\theta}^{(-k)} = \boldsymbol{\theta} \setminus \theta_k = \{\theta_1, \ldots, \theta_m\} - \theta_k$, $\gamma'$ is an increasing function of $\theta_k$.*

A proof of lemma 1 is provided in Appendix A.1.

Given that the right-hand side of (7) is an increasing function of $\theta_k$, the maximum of the upper bound of $\gamma$ can be obtained by condition (7) on the least favorable configuration of $H_0$ which, as Tamhane and Logan (2004) noted, is either $\cap_{k=1}^{m}(\theta_k = \epsilon_k)$ or $(\theta_k = -\eta_k) \cap (\cap_{i=1}^{m \neq k}(\theta_i \to \infty))$. The following theorem provides explicit formulas for the determination of the critical value $t_{d,a'}$ and hence the adjusted significance level $\alpha'$ that maximize the upper bound of the Type I error probability under the least favorable configuration of the combined superiority and non-inferiority hypotheses.

**Theorem 1.** *Let $\gamma_1(\alpha')$ and $\gamma_2(\alpha')$ be defined as*

$$\gamma_1(\alpha') = \sum_{k=1}^{m} P_0\{(T_k > t_{d,a'}) \cap (\cap_{i=1}^{m, i \neq k}(T_i > t_{d,a'} - c_k))\} \tag{8}$$

$$\gamma_2(\alpha') = \max_{1 \leq k \leq m} P_0\{T_k > t_{d,a'} + c_k\} + (m-1)P_0\{T_k > t_{d,a'}\}. \tag{9}$$

*Then the overall Type I error rate (6) is controlled at $\alpha$ if $\alpha'$ is chosen such that*

$$\max\left\{\gamma_1(\alpha'), \gamma_2(\alpha')\right\} \leq \alpha. \tag{10}$$

A proof of Theorem 1 can be found in Appendix A.2.

The result in (10) provides the basis of finding the significance level $\alpha'$ that is appropriately adjusted for multiple tests such that the overall Type I error is controlled at $\alpha$. When the $T_k$'s are exchangeable, that is, their probability distribution functions are symmetric in their arguments (e.g., $t_{d,a'}$, $\rho_{k_1 k_2}$ and $c_k$), the quantities $\gamma_1(\alpha')$ in (8) and $\gamma_2(\alpha')$ (9) reduce to

$$\gamma_1(\alpha') = mP_0\{(T_1 > t_{d,a'}) \cap (\cap^{m}_{i=2}(T_i > t_{d,a'} - c))\}, \tag{11}$$

$$\gamma_2(\alpha') = P_0\{T_1 > t_{d,a'} + c\} + (m-1)P\{T_1 > t_{d,a'}\} \tag{12}$$

with $c = c_1 = \ldots = c_m$.

### 3.2 Critical value $t_{d,a'}$

The determination of $t_{d,a'}$ or $\alpha'$ is the key for the simultaneous tests of superiority and non-inferiority hypotheses. While the algorithm of searching for $\alpha'$ is readily available, further explorations of $\alpha'$ may help elucidate its relationship with data structure and the pre-specified superiority and non-inferiority effect margins.

In order to determine $\alpha'$ and hence $t_{d,a'}$ that satisfies (10), first note that given the input $m$, $\rho_{k_1 k_2}$, and $d$, both the error probabilities $\gamma_1(\alpha')$ and $\gamma_2(\alpha')$ are increasing functions of

$\alpha'$. In addition,

$$\gamma_1(\alpha') = mP_0\{\cap_{k=1}^{m}(T_k > t_{d,a'})\}$$
$$\leq \gamma_2(\alpha') = mP_0\{T_k > t_{d,a'}\} \tag{13}$$

as $c_k \to 0$ for all $k$'s and

$$\gamma_1(\alpha') = \sum_{k=1}^{m} P_0\{T_k > t_{d,a'}\}$$
$$\geq \gamma_2(\alpha') = (m-1)P_0\{T_k > t_{d,a'}\} \tag{14}$$

as $c_k \to \infty$ for all $k$'s. Also note that $\alpha' = \alpha$ under certain extreme cases such as perfect correlation among all endpoints. Hence, the largest and smallest possible values of $\alpha'$ are $\alpha$ and $\alpha/m$, respectively, with the later being the Bonferroni-adjusted significance level for each test.

To search for $\alpha'$, the algorithm of bisection method of Gentle (2009, pp.246-248) is adopted. First let $\alpha'_{l(0)} = \alpha/m$ and $\alpha'_{u(0)} = \alpha$. After choosing $\alpha$ (say, $\alpha = 0.05$ or $0.025$ for one-sided) and the precision $\xi \geq 0$ (say, $\xi = 10^{-5}$), one follows the steps described below to find an appropriate $\alpha'$ value that satisfies (10):

1. Set $s = 0$ and calculate $\alpha'_{(0)} = \left(\alpha'_{l(0)} + \alpha'_{u(0)}\right)/2$;

2. Calculate $\Delta_{(0)} = \alpha'_{u(0)} - \max\left[\gamma_1(\alpha'_{(0)}), \gamma_2(\alpha'_{(0)})\right]$. If $0 < \Delta_{(0)} \leq \xi$ then stop and return $\alpha'_{(0)}$ as the solution of $\alpha'$; otherwise let $\alpha'_{l(1)} = \alpha'_{(0)}$ and $\alpha'_{u(1)} = \alpha'_{u(0)}$;

3. Calculate $\alpha'_{(1)} = \left(\alpha'_{l(1)} + \alpha'_{u(1)}\right)/2$ and $\Delta_{(1)} = \alpha'_{u(1)} - \max\left[\gamma_1(\alpha'_{(1)}), \gamma_2(\alpha'_{(1)})\right]$. If $\Delta_{(1)} \leq \xi$, then stop and return $\alpha'_{(1)}$ as the solution of $\alpha'$.

4. In general, at step $s > 1$ if $0 < \Delta_{(s)} \leq \xi$ then stop and return $\alpha'_{(s)}$ as the solution of $\alpha'$; otherwise let $\alpha'_{l(s)} = \alpha'_{(s-1)}$ and $\alpha'_{u(s)} = \alpha'_{u(s-1)}$;

5. Calculate $\alpha'_{(s)} =$ and $\Delta_{(s)} = \alpha'_{u(s)} - \max[\gamma_1(\alpha'_{(s)}), \gamma_2(\alpha'_{(s)})]$ and compare $\Delta_{(s)}$ with $\xi$ and return to Step 4.

The computation of the error probabilities (8) and (9) requires the correlation matrix or partial correlation coefficients $\rho_{k_1 k_2}$ of the test statistics. While one can always plug in the estimates of the $\rho_{k_1 k_2}$'s in the formula, an alternative yet simple approach when the $\rho_{k_1 k_2}$'s are approximately equal is to estimate the mean correlation coefficient $\rho_{k_1 k_2}$ by the method of Armitage and Parmar (1986)

$$\rho_0 = \frac{2\sum_{i=1}^{m-1}\sum_{j=i+1}^{m}|\rho_{ij}|}{m(m-1)} + \frac{4\sum_{i=1}^{m-1}\sum_{j=i+1}^{m}(|\rho_{ij}| - \frac{2\sum_{i=1}^{m-1}\sum_{j=i+1}^{m}|\rho_{ij}|}{m(m-1)})^2}{m(m-1)} \tag{15}$$

and then use $\rho_0$ as the common partial correlation coefficient in (8) and (9) or (11) and (12) to obtain the error probabilities.

Given the number of endpoints $m$, common partial correlation coefficient $\rho_0$, degrees of freedom $d$, and the common combined standardized margins $c$ for superiority and non-inferiority, one can directly calculate the critical values $t_{d,\alpha'}$ using the above bisection method. Figure 1 shows how the the critical values $t_{d,\alpha'}$ change over the domain of combined standardized margin $c$ from 0 to 5 (exchangeable case) for $m = (2, 3)$, $\rho = (0, 0.5)$, $d = (10, 50, 200)$. The behavior of $t_{d,\alpha'}$ can be summarized as follows:

(a) There is no monotonic trend in critical value $t_{d,\alpha'}$ over effect margin. As $c$ increases, $t_{d,\alpha'}$ decreases initially, then reaches the lowest point at certain $c$ value after which it starts increasing. This is because the adjusted $\alpha'$ is determined by $\gamma_1$ when $c$ is large as implied in (14) and by $\gamma_2$ when $c$ is small as indicated in (13). The turning point of $c$ depends on $m$, $\rho$ and $d$.

(b) Given the effect margin $c$, the critical value $t_{d,\alpha'}$ increases with $m$ and $\rho$.

*(c) The impact of degree of freedom $d$ on $t_{d,\alpha'}$ becomes smaller for sufficiently larger $d$ values.*

Given a critical value $t_{d,\alpha'}$, the adjusted $\alpha'$ can be calculated conversely. Table 1 shows the adjusted $\alpha'$ values for selected combinations of $\rho$, $c$ and $d$ for $m = 2, 3$. The key message for Table 1 can be summarized as follows: (i) As the common effect margin $c$ increases, the adjusted $\alpha'$ increases and then decreases, following the inverse pattern of $t_{d,\alpha'}$ over the same range of $c$ value, (ii) when $c = 0$ or $c$ is large enough, $\alpha'$ approaches $\alpha/m$ regardless of $\rho$ or $d$, which can be explained by (13) and (14), and (iii) the adjusted $\alpha'$ decreases as $m$ and/or $\rho$ increases.

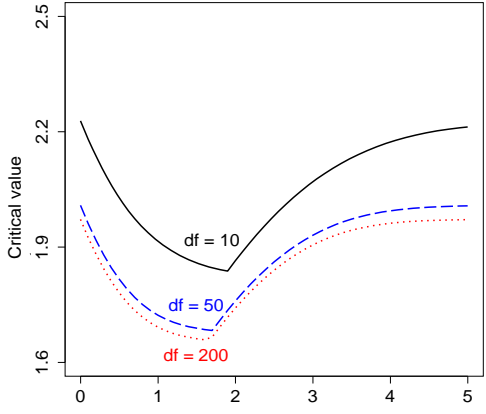

(a) $m = 2$ and $\rho = 0$

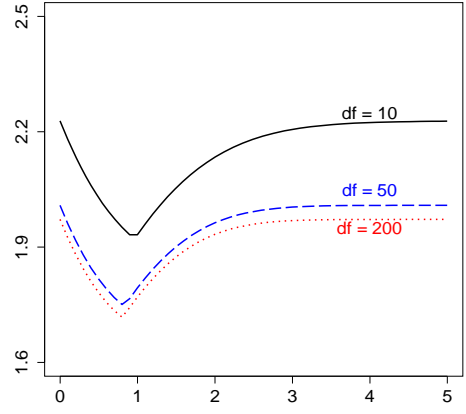

(b) $m = 2$ and $\rho = 0.5$

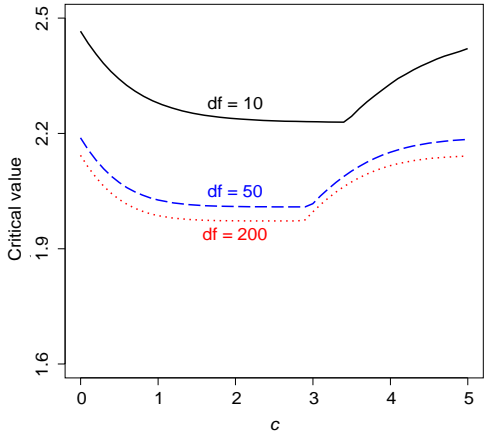

(c) $m = 3$ and $\rho = 0$

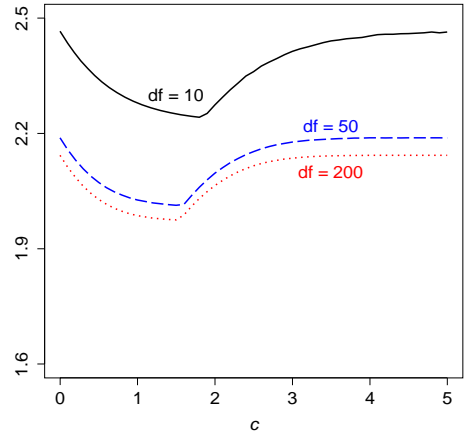

(d) $m = 3$ and $\rho = 0.5$

Figure 1: Critical values $t_{d,a'}$ as a function of the number of endpoints $m$, common correlation coefficient $\rho$, degrees of freedom $d$ and a common effect margin $c$.

Once $\alpha'$ and hence $t_{d,a'}$ are determined, one can easily calculate the exact $(1 - \alpha)\%$ simultaneous one-sided confidence intervals $[L_1, \infty) \times \ldots \times [L_k, \infty) \times \ldots \times \ldots [L_K, \infty)$, where $L_k$ is defined as

$$L_k = \hat{\theta}_k - t_{d,a'} \hat{\sigma}_k \sqrt{1/n_1 + n_2}$$

Table 1: Adjusted $\alpha'$ values for $m = 2, 3$ and various choices of $\rho$, $c$ and $d$

| | | | \multicolumn{7}{c}{Degrees of freedom $d$} | | | | | | |
| --- | --- | --- | --- | --- | --- | --- | --- | --- | --- |
| $m$ | $\rho$ | $c$ | 10 | 20 | 30 | 40 | 50 | 100 | 200 |
| 2 | 0.0 | 0.0 | 0.0250 | 0.0250 | 0.0250 | 0.0250 | 0.0250 | 0.0250 | 0.0250 |
| | | 0.5 | 0.0350 | 0.0366 | 0.0372 | 0.0375 | 0.0377 | 0.0380 | 0.0382 |
| | | 1.0 | 0.0423 | 0.0443 | 0.0450 | 0.0454 | 0.0456 | 0.0459 | 0.0461 |
| | | 2.0 | 0.0460 | 0.0437 | 0.0429 | 0.0425 | 0.0423 | 0.0418 | 0.0416 |
| | | 3.0 | 0.0327 | 0.0307 | 0.0301 | 0.0298 | 0.0296 | 0.0292 | 0.0291 |
| | | 4.0 | 0.0274 | 0.0263 | 0.0260 | 0.0259 | 0.0258 | 0.0256 | 0.0256 |
| | | 5.0 | 0.0257 | 0.0252 | 0.0251 | 0.0251 | 0.0251 | 0.0250 | 0.0250 |
| | 0.5 | 0.0 | 0.0250 | 0.0250 | 0.0250 | 0.0250 | 0.0250 | 0.0250 | 0.0250 |
| | | 0.5 | 0.0350 | 0.0366 | 0.0372 | 0.0375 | 0.0377 | 0.0380 | 0.0382 |
| | | 1.0 | 0.0411 | 0.0401 | 0.0398 | 0.0396 | 0.0395 | 0.0392 | 0.0391 |
| | | 2.0 | 0.0293 | 0.0282 | 0.0279 | 0.0277 | 0.0276 | 0.0274 | 0.0273 |
| | | 3.0 | 0.0260 | 0.0255 | 0.0253 | 0.0253 | 0.0253 | 0.0252 | 0.0252 |
| | | 4.0 | 0.0252 | 0.0250 | 0.0250 | 0.0250 | 0.0250 | 0.0250 | 0.0250 |
| | | 5.0 | 0.0250 | 0.0250 | 0.0250 | 0.0250 | 0.0250 | 0.0250 | 0.0250 |
| 3 | 0.0 | 0.0 | 0.0167 | 0.0167 | 0.0167 | 0.0167 | 0.0167 | 0.0167 | 0.0167 |
| | | 0.5 | 0.0206 | 0.0213 | 0.0215 | 0.0216 | 0.0217 | 0.0219 | 0.0219 |
| | | 1.0 | 0.0229 | 0.0236 | 0.0239 | 0.0240 | 0.0240 | 0.0241 | 0.0242 |
| | | 2.0 | 0.0246 | 0.0249 | 0.0249 | 0.0249 | 0.0249 | 0.0250 | 0.0250 |
| | | 3.0 | 0.0249 | 0.0250 | 0.0250 | 0.0248 | 0.0245 | 0.0240 | 0.0237 |
| | | 4.0 | 0.0211 | 0.0191 | 0.0186 | 0.0183 | 0.0182 | 0.0179 | 0.0178 |
| | | 5.0 | 0.0180 | 0.0171 | 0.0170 | 0.0169 | 0.0168 | 0.0168 | 0.0168 |
| | 0.5 | 0.0 | 0.0167 | 0.0167 | 0.0167 | 0.0167 | 0.0167 | 0.0167 | 0.0167 |
| | | 0.5 | 0.0206 | 0.0213 | 0.0215 | 0.0216 | 0.0217 | 0.0219 | 0.0219 |
| | | 1.0 | 0.0229 | 0.0236 | 0.0239 | 0.0240 | 0.0240 | 0.0241 | 0.0242 |
| | | 2.0 | 0.0230 | 0.0215 | 0.0210 | 0.0208 | 0.0206 | 0.0203 | 0.0201 |
| | | 3.0 | 0.0183 | 0.0175 | 0.0172 | 0.0172 | 0.0171 | 0.0170 | 0.0170 |
| | | 4.0 | 0.0170 | 0.0168 | 0.0167 | 0.0167 | 0.0167 | 0.0167 | 0.0167 |
| | | 5.0 | 0.0167 | 0.0167 | 0.0167 | 0.0167 | 0.0167 | 0.0167 | 0.0167 |

for a two-sample problem with estimated mean treatment effect $\hat{\theta}_k$, pooled standard deviation $\hat{\sigma}_k$ and sample sizes $n_1$ and $n_2$ for respective groups.

## 4 Type I error probability when $p = r$

Suppose that a clinical trial is considered positive if and only if superiority is demonstrated on at least $p$ out of the $m$ endpoints and noninferiority on the rest $m - p$ endpoints. Then, the type I error probability of rejecting at least $p$ out of $m$ $H_{0k}^{S}$'s and the rest $m - p$ $H_{0k}^{NI}$'s is given by

$$\gamma_p = P_0 \left\{ \bigcup_{v=p}^{m} \bigcap_{i=1}^{v} \left( T_{k_i} > t_{d,a'} - \eta'_{k_i}(\theta_{k_i}) + c_{k_i} \right) \cap \bigcap_{j=1, j \neq k_1,\ldots,k_v}^{m} T_{k_j} > t_{d,a'} - \eta'_{k_j}(\theta_{k_j}) \right\}. \quad (16)$$

Define the usual Bonferroni summands

$$S_{v,m} = \sum_{1 \leq k_1 < \ldots < k_v \leq m} P_0 \left\{ \bigcap_{i=1}^{v} \left( T_{k_i} > t_{d,a'} - \eta'_{k_i}(\theta_{k_i}) + c_{k_i} \right) \cap \bigcap_{j=1, j \neq k_1,\ldots,k_v}^{m} T_{k_j} > t_{d,a'} - \eta'_{k_j}(\theta_{k_j}) \right\} \quad (17)$$

with the summation taken over all subscripts satisfying $1 \leq k_1 < \ldots < k_v \leq m$. Then, the type I error probability (16) can be written as

$$\gamma_p = \sum_{v=p}^{m} (-1)^{v-p} \binom{v-1}{p-1} S_{v,m}, \quad (18)$$

see, e.g., Feller (1968, p.109). A sharper upper bound on $\gamma_p$, due to Hoppe and Seneta (1990), is given by

$$\gamma_p \leq S_{p,m} - \sum_{a \leq k_1 < \ldots < k_v \leq m} P_0 \left\{ \left( T_k > t_{d,a'} - \eta'_k(\theta_k) + c_k \right) \cap \bigcap_{i=1}^{v} \left( T_{k_i} > t_{d,a'} - \eta'_{k_i}(\theta_{k_i}) + c_{k_i} \right) \cap \bigcap_{j=1, j \neq k_1,\ldots,k_v, k}^{m} T_{k_j} > t_{d,a'} - \eta'_{k_j}(\theta_{k_j}) \right\} \quad (19)$$

where the summation is taken over all $k \notin \{k_1, \ldots, k_v\}$ and $k \leq k_v - 1$. Note that per the definition of $S_{v,m}$ in (17), the right-hand side of (19) can also be written as

$$\gamma_p \leq \sum_{a \leq k_1 < \ldots < k_v \leq m} P_0 \left\{ \left( t_{d,a'} - \eta'_k(\theta_k) < T_k < t_{d,a'} - \eta'_k(\theta_k) + c_k \right) \cap \bigcap_{i=1}^{v} T_{k_i} > t_{d,a'} - \eta'_{k_i}(\theta_{k_i}) + c_{k_i} \cap \bigcap_{j=1, j \neq k_1,\ldots,k_v, k}^{m} T_{k_j} > t_{d,a'} - \eta'_{k_j}(\theta_{k_j}) \right\} \quad (20)$$

which is clearly increasing in $\{\eta'_{k_i}(\theta_{k_i}) : 1 \leq k_1 < \ldots < k_v \leq m\}$ for all $k \notin \{k_1, \ldots, k_v\}$.

**Theorem 2.** *Let $\gamma_1(\alpha')$ and $\gamma_2(\alpha')$ be defined as*

$$\gamma_1(\alpha'_p) = \max_{1 \leq k_1 < \ldots < k_p \leq m} P_0\{\cap_{i=1}^{p}(T_{k_i} > t_{d,a'}) \cap (\cap_{i=1}^{m \neq k_1,\ldots,k_p}(T_i > t_{d,a'} - c_k))\} \quad (21)$$

$$\gamma_2(\alpha'_p) = \frac{m}{p} \max_{1 \leq k \leq m} P_0\{T_k > t_{d,a'} + c_k\} + \frac{m}{m-p} P_0\{T_k > t_{d,a'}\}. \quad (22)$$

*Then the Type I error rate (16) is controlled at $\alpha$ if $\alpha'$ is chosen such that*

$$\max\left\{\gamma_1(\alpha'), \gamma_2(\alpha')\right\} \leq \alpha. \quad (23)$$

*Proof*: It follows the proof of Theorem **??** and the rejection probability equation (16).

When the test statistics $T_k$'s are exchangeable with respect to their probability distributional functions, the quantities $\gamma_1(\alpha'_p)$ in (21) and $\gamma_2(\alpha'_p)$ in (22) reduce to

$$\gamma_1(\alpha'_p) = \frac{m}{p} P_0\left\{\cap_{k=1}^{p}(T_k > t_{d,a'}) \cap (\cap_{i=p+1}^{m}(T_i > t_{d,a'} - c))\right\}; \quad (24)$$

$$\gamma_2(\alpha'_p) = \frac{m}{p} P_0\left\{T_k > t_{d,a'} + c\right\} + \frac{m}{m-p} P_0\left\{T_k > t_{d,a'}\right\}. \quad (25)$$

## 5 Simulations

Simulations are performed to compare the overall Type I error rate and power of our proposed approach (abbreviated as CCZQ hereafter) with three existing methods whose algorithms are briefly described below.

Tamhane and Logan (2004)'s UI-IU test (referred to as TL hereafter) uses bootstrap algorithm to obtain sharper critical values in a stepwise manner. Specifically, the critical values $d_3$ and $d_4$ are solutions in the probability equation

$$P_0\left\{\min_{1 \leq k \leq m}(T_k + c_k) > d_3 \text{ and } \max_{1 \leq k \leq m} T_k > d_4\right\} = \alpha \quad (26)$$

where $d_3$ is set to equal $t_{d,a}$ and $T_1, \ldots, T_m$ have the multivariate $t$-distribution. In Bloch et al. (2001)'s bootstrap-based method (referred to as BLT hereafter), the superiority null hypothesis $H_0^S$ is tested using one-sided Hotelling $T^2$ test statistic which is modified to allow for unequal covariance matrices, and the non-inferiority null hypothesis $H_0^{NI}$ is tested using the IU test with rejection region given by $\min_{1 \leq k \leq m} t^{NI}_k > t_{d,a}$. The superiority and non-inferiority null hypothesis $H_0$ is rejected if

$$T^2 \times I\left(\min_{1 \leq k \leq m} t^{NI}_k > t_{d,a}\right) > d_1, \tag{27}$$

where $I$ denotes an indicator function, $d_1 > 0$ is a critical value that is determined by the Type I error probability. Perlman and Wu (2004) (referred to as PW hereafter) replace the Hotelling $T^2$ statistic in Bloch et al. (2001) by the multivariate one-sided likelihood ratio test statistic derived by Perlman (1969), and test $H^S_0$ and $H^{NI}_0$ separately, each at level $\alpha$, with the latter using the IU test. The test statistic for testing $H^S$, denoted by $U^2$, is the difference between the observed vector $\bar{x}_1 - \bar{x}_2$ and its projection on to the non-positive orthant $O^- = \{\theta | \theta_k \leq 0 \text{ for all } k\}$. Then $H_0$ is rejected if $U^2 > d_2$ and $\min_{1 \leq k \leq m} t^{NI}_k > t_{d,a}$, where $d_2$ is the solution to

$$\frac{1}{2} P\left(\frac{\chi^2_{m-1}}{\chi^2_{n_1+n_2-m}} > d_2\right) + \frac{1}{2} P\left(\frac{\chi^2_m}{\chi^2_{n_1+n_2-m-1}} > d_2\right) = \alpha. \tag{28}$$

We consider simulation settings that are similar to those in Logan and Tamhane (2008), that is, two-arm studies with a sample size $n_1 = n_2 = 100$, $m = 2$ endpoints, correlation coefficient $\rho = (0, 0.5)$ and the combined common standardized superiority and non-inferiority effect margins $c = (0.2, 0.33, 0.5)$. All results are based on 10,000 simulations. Table 2 presents the overall Type I error rates which are controlled at a pre-defined level $\alpha = 0.05$ for all the methods in comparison (except for TL method with $\rho = 0.5$ and $c = 0.33$ for which the estimated Type I error rate = 0.051 that is considered within the simulation error). It is noted that the estimates of Type I error rates for our proposed approach are closest to the pre-defined nominal level 0.05 among all the methods in comparison, regardless of the correlation coefficient $\rho$ and the common standardized superiority and non-inferiority effect margins $c$.

Simulations for power comparisons are performed using the same simulation setting as in the Type I error simulation with the following pairs of treatment effects on the two endpoints: $(\theta_1, \theta_2) = (0.4, 0), (0.66, 0), (0.4, 0.2)$ and $(0.33, 0.33)$. Table 3 presents the results of power simulation for our approach and the three existing methods. Obviously, our unified approach outperforms all other three methods in terms of power for all scenarios under consideration. In fact, after determining $\alpha'$ and hence $t_{d,a'}$ from (10), the power for our proposed approach can be directly calculated using the formula

$$\text{Power} = P_1\{\cap_{k=1}^{m}(T_k > t_{d,a'} - \eta'_k(\theta_k^1))\} \\ - P_1\{(\cap_{k=1}^{m}(t_{d,a'} - \eta'_k(\theta_k^1) < T_k \leq t_{d,a'} - \eta'(\theta_k^1) + c_k))\} \quad (29)$$

where $\eta'_k(\theta_k^1)$ is as defined in Section 2 with $\theta_k$ being replaced by $\theta_k^1$, an assumed value of $\theta_k$ under the alternative hypothesis for which the power is calculated. Equation (29) can also be used to determine the minimum sample size for which a desired power can be achieved.

Table 2: Simulated Type I error probabilities of our proposed approach and three existing methods with pre-defined $\alpha = 0.05$

| $\rho$ | $c$ | $(\theta_1, \theta_2)$ | CCZQ | TL | PW | BLT |
|---|---|---|---|---|---|---|
| 0 | 0.2 | (0,0) | 0.050 | 0.016 | 0.022 | 0.023 |
|  | 0.33 | (0,0) | 0.050 | 0.031 | 0.032 | 0.033 |
|  | 0.5 | (0,0) | 0.048 | 0.044 | 0.042 | 0.032 |
| 0.5 | 0.2 | (0,0) | 0.048 | 0.034 | 0.037 | 0.039 |
|  | 0.33 | (0,0) | 0.050 | 0.051 | 0.049 | 0.043 |
|  | 0.5 | (0,0) | 0.050 | 0.050 | 0.050 | 0.044 |

## 6 Examples

Two examples are provided below to illustrate the use of the proposed approach, one with two primary endpoints and the other with four primary endpoints. These two examples are also discussed in Logan and Tamhane (2008).

### 6.1 Example 1

Röhmel et al. (2006) illustrate an example in a confirmatory clinical trial comparing a treatment with a placebo control (note that in practice an active control should be used to demonstrate non-inferiority of the new treatment over the active control which should retain some therapeutic effect than a placebo control). The trial randomized $n_1 = 442$ patients to the treatment group and $n_2 = 211$ patients to the control group. Two primary endpoints taken from each patient were assumed to follow approximately bivariate normal distribution with low values of the endpoints indicating beneficial effects. The summary statistics of the endpoints are as follows

$$\begin{pmatrix} \bar{x}_{1t} \\ \bar{x}_{2t} \end{pmatrix} \sim N\left(\begin{pmatrix} 13.269 \\ 22.796 \end{pmatrix}, \begin{pmatrix} 78.60082 & 36.12524 \\ 36.12524 & 111.65005 \end{pmatrix}\right),$$

$$\begin{pmatrix} \bar{x}_{1c} \\ \bar{x}_{2c} \end{pmatrix} \sim N\left(\begin{pmatrix} 15.322 \\ 23.512 \end{pmatrix}, \begin{pmatrix} 100.13374 & 53.62950 \\ 53.62950 & 130.84153 \end{pmatrix}\right),$$

where $\bar{x}_{1t}$ and $\bar{x}_{2t}$ denote respectively the sample means of the first and second endpoints for treatment group, and $\bar{x}_{1c}$ and $\bar{x}_{2c}$ the sample means of the first and second endpoints for the control group. Röhmel et al. (2006) apply Läuter (1996)'s standard sum (SS) test for simultaneously testing superiority on at least one endpoint with superiority margins $\epsilon_1 = \epsilon_2 = 0$ and non-inferiority on both endpoints with non-inferiority margins $\eta_1 = 1$ and $\eta_2 = 2$. They obtain the global Läuter's SS test statistic 2.0416 which exceeds the critical value 1.9636 at one-sided 0.025 significance level, leading to the rejection of the superiority null hypothesis. They then compare the upper limit of the two-sided 95% confidence interval on the mean difference of each endpoint to its corresponding non-inferiority margin and conclude that the treatment is non-inferior over the control on both endpoints, which, together with the superiority test, leads to the conclusion that the trial successfully meets its objective. Logan and Tamhane (2008) also use this example data in their stepwise testing procedure in which they first calculate the $t$ test statistics $t_1^{NI} = 3.945$ and $t_2^{NI} = 2.990$ for non-inferiority and compare these $t$ statistics with the critical value 1.96, which leads to the rejection of non-inferiority hypotheses. They then apply a global test for testing superiority null hypothesis by comparing the maximum superiority test statistic $t^S_{max} = 2.653$ to the critical value 2.114 (obtained from bootstrap accounting for rejection of non-inferiority null hypotheses) or 2.220 (the upper 0.0215 percentile of standard normal distribution that does not take into account the rejection of non-inferiority null hypotheses) and conclude that the treatment is superior over the control on the fist endpoint.

Table 3: Simulated powers of our proposed approach and three existing methods with $\epsilon=0$, and $\alpha =0.05$

| $\rho$ | $\eta$ | $(\theta_1, \theta_2)$ | CCZQ | TL | PW | BLT |
|---|---|---|---|---|---|---|
| 0 | 0.2 | (0.4,0) | 0.269 | 0.214 | 0.227 | 0.239 |
| | | (0.66,0) | 0.319 | 0.306 | 0.304 | 0.291 |
| | | (0.4,0.2) | 0.737 | 0.630 | 0.681 | 0.696 |
| | | (0.33,0.33) | 0.851 | 0.742 | 0.810 | 0.817 |
| | 0.3 | (0.4,0) | 0.585 | 0.466 | 0.468 | 0.433 |
| | | (0.66,0) | 0.689 | 0.640 | 0.644 | 0.645 |
| | | (0.4,0.2) | 0.869 | 0.746 | 0.796 | 0.790 |
| | | (0.33,0.33) | 0.901 | 0.778 | 0.842 | 0.844 |
| | 0.5 | (0.4,0) | 0.823 | 0.683 | 0.670 | 0.489 |
| | | (0.66,0) | 0.957 | 0.932 | 0.929 | 0.862 |
| | | (0.4,0.2) | 0.904 | 0.770 | 0.812 | 0.798 |
| | | (0.33,0.33) | 0.915 | 0.777 | 0.845 | 0.849 |
| 0.5 | 0.2 | (0.4,0) | 0.305 | 0.268 | 0.266 | 0.255 |
| | | (0.66,0) | 0.320 | 0.304 | 0.301 | 0.295 |
| | | (0.4,0.2) | 0.740 | 0.645 | 0.650 | 0.642 |
| | | (0.33,0.33) | 0.808 | 0.701 | 0.723 | 0.743 |
| | 0.3 | (0.4,0) | 0.629 | 0.535 | 0.529 | 0.356 |
| | | (0.66,0) | 0.690 | 0.650 | 0.647 | 0.611 |
| | | (0.4,0.2) | 0.842 | 0.734 | 0.738 | 0.683 |
| | | (0.33,0.33) | 0.835 | 0.706 | 0.725 | 0.752 |
| | 0.5 | (0.4,0) | 0.832 | 0.710 | 0.690 | 0.364 |
| | | (0.66,0) | 0.958 | 0.938 | 0.932 | 0.755 |
| | | (0.4,0.2) | 0.866 | 0.740 | 0.744 | 0.689 |
| | | (0.33,0.33) | 0.850 | 0.719 | 0.735 | 0.746 |

We now apply our method to the above data by first calculating the standard $t$ test statistics $t_1 = (15.322 - 13.269)/(100.13374/211 + 78.60082/442)^{\frac{1}{2}} = 2.5418$ and $t_2 = (23.512 - 22.796)/(130.84153/211 + 111.65005/442)^{\frac{1}{2}} = 0.7664$, with correlation coefficient $\rho = (36.12524 + 53.6295)/[(78.60082 + 100.13374)(111.65005 + 130.84153)]^{\frac{1}{2}} = 0.4311$. The standardized non-inferiority margins are $c_1 = 1/(78.60082/442 + 100.13374/211)^{\frac{1}{2}} = 1.2380$ and $c_2 = 2/(111.65005/442 + 130.84153/211)^{\frac{1}{2}} = 2.1409$. Plugging these numerical values into (8) and (9), we obtain $\alpha' = 0.0243$ from (10) which controls the overall Type I error rate at $\alpha = 0.025$, and hence have $t_{651, 0.0243} = 1.9758$. Since $t_1 > 1.9758$ and $t_2 + c_2 > 1.9758$, we conclude that the treatment is superior over the control on the first endpoint and non-inferior on the second endpoint. The exact 97.5% simultaneous one-sided confidence intervals on the mean difference of treatment effects on both endpoints are given by

$$[15.322 - 13.269 - t_{651,0.0243}(100.13374/211 + 78.60082/442)^{\frac{1}{2}}, \infty)$$
$$\times [23.512 - 22.796 - t_{651,0.0243}(130.84153/211 + 111.65005/442)^{\frac{1}{2}}, \infty)$$

which reduces to $[0.4571, \infty) \times [-1.1298, \infty)$.

## 6.2 Example 2

Zhang et al. (1997) present a randomized clinical trial comparing a new drug ($n_1 = 34$) with a control ($n_2 = 35$) in terms of efficacy and safety among asthmatic patients. The trial includes the following four primary endpoints

1. forced expiratory volume in 1 second ($FEV_1$), in liters;

2. Symptoms score (SS), 0-6 scale;

3. Peak expiratory flow rate (PEFR), in liters per minute;

4. Additional medication use (AMU), i.e., agonist use, in puffs per day.

Sample means by group and pooled standard deviations for the four endpoints are given in Table 4. The estimated partial correlation coefficients $\rho_{k_1 k_2}$ are presented in Table 5.

Table 4: The univariate analysis for the data of the asthma example

| Group | Statistics | $FEV_1$ | SS | PEFR | AMU |
|---|---|---|---|---|---|
| Test drug | $\bar{x}_{kt}$ | 14.0 | 0.86 | 16.5 | 0.49 |
| Control | $\bar{x}_{kc}$ | 5.7 | 0.34 | 1.6 | 0.15 |
| Pooled SD ($\hat{\sigma}_k$) | | 11.5 | 0.96 | 22.3 | 0.66 |

Table 5: The correlation among the four endpoints

| | $PEV_1$ | SS | PEFR |
|---|---|---|---|
| SS | 0.31 | | |
| PEFR | 0.25 | 0.42 | |
| AMU | 0.24 | 0.67 | 0.43 |

Before applying our approach to this data, we first calculate the common correlation coefficient $\rho_0 = 0.4298$ by equation (15) and then use formulas (8)-(10) to determine $\alpha' =$ 0.01254 that controls the overall Type I error rate at $\alpha = 0.025$ (one-sided), which yields $t_{67, 0.01254} = 2.2917$ (If we use the original correlation coefficients in Table (5), we obtain an adjust significance level $\alpha' = 0.01274$ and the corresponding critical value $t_{67, 0.01274} = 2.2850$). Now we set a common superiority margin $\epsilon = 0$ and the standardized non-inferiority margins $c_k = 0.2\sigma_k$ for $k = 1, 2, 3, 4$, where $\sigma_k$ denotes the pooled standard deviation for the $k$th endpoints (These non-inferiority margins are also used in Logan and Tamhane (2008)). Based on the summary statistics in Table (4), we obtain the $t$ test statistics $t_1 = 2.9973$, $t_2 = 2.2495$, $t_3 = 2.7748$ and $t_4 = 2.1394$ for $FEV_1$, SS, PEFR and AMU, respectively. Using these $t$ test statistics and the non-inferiority margins, we can conclude that the new drug is non-inferior over the control for all the four endpoints, and superior in $FEV_1$ and PEFR. The exact 97.5% simultaneous one-sided confidence intervals are therefore given by $[1.9539, \infty) \times [-0.0098, \infty) \times [2.59419, \infty) \times [-0.0242, \infty)$.

## 7 Discussions and Conclusions

We have proposed a unified approach to simultaneous testing of superiority and non-inferiority hypotheses on multiple endpoints that are commonly seen in clinical trials. The proposed approach is based on the UI-IU test of Tamhane and Logan (2004) and the least favorable configurations of the superiority and non-inferiority null hypotheses, which leads to the solution of adjusted significance level $\alpha'$ for marginal tests that controls the overall Type I error rate at pre-defined $\alpha$. Unlike existing methods which test superiority and non-inferiority null hypotheses separately and control the Type I error rate each at $\alpha$, our method provides a unified solution for testing superiority and non-inferiority hypotheses simultaneously using the derived significance level $\alpha'$ and its corresponding critical value that depend not only on the number of endpoints, correlation coefficients, and sample sizes, but also on the combined standardized superiority and non-inferiority margins. Simulation studies show that our proposed approach maintains a higher power than other available methods in scenarios under investigation. Since the adjusted significance level $\alpha'$ is derived that controls the overall Type I error rate at $\alpha$, the $(1 - \alpha)$% simultaneous confidence intervals can be constructed, which is obviously another advantage of our approach.

The proposed method is based on general probability distributions and uses multivariate $t$ distribution of continuous data as an illustration. Therefore, with appropriate modifica-tion, the method can be extended to discrete outcomes and survival data. Some further improvements are under development to include sequential testing of superiority and non-inferiority hypotheses as well as simultaneous testing of superiority on at least $m_1 \geq 2$ endpoints and non-inferiority on the remaining $m - m_1$ endpoints.

# A Appendix

## A.1 Proof of lemma 1

*Proof*: Note that for any arbitrary $k_1$, $\gamma'$ can be written as

$$
\begin{aligned}
\gamma' &= P_0 \left\{ [T_{k_1} > t_{d,\alpha'} - \eta'_{k_1}(\theta_{k_1}) + c_{k_1}] \cap \left[ \cap_{i=1}^{m, \neq k_1} (T_i > t_{d,\alpha'} - \eta'_i(\theta_i)) \right] \right\} \\
&+ \sum_{k=1}^{m, \neq k_1} \left[ P_0 \left\{ [T_k > t_{d,\alpha'} - \eta'_k(\theta_k) + c_k] \cap \left[ \cap_{i=1}^{m, \neq k} (T_i > t_{d,\alpha'} - \eta'_i(\theta_i)) \right] \right\} \right. \\
&\quad - P_0 \left\{ (T_{k_1} > t_{d,\alpha'} - \eta'_{k_1}(\theta_{k_1}) + c_{k_1}) \cap (T_k > t_{d,\alpha'} - \eta'_k(\theta_k) + c_k) \right. \\
&\quad \left. \left. \cap \left[ \cap_{j=1}^{m, \neq k_1, k} (T_j > t_{d,\alpha'} - \eta'_j(\theta_j)) \right] \right\} \right] \\
&= P_0 \left\{ [T_{k_1} > t_{d,\alpha'} - \eta'_{k_1}(\theta_{k_1}) + c_{k_1}] \cap \left[ \cap_{i=1}^{m, \neq k_1} (T_i > t_{d,\alpha'} - \eta'_i(\theta_i)) \right] \right\} \\
&+ \sum_{k=1}^{m, \neq k_1} P_0 \left\{ [T_{k_1} > t_{d,\alpha'} - \eta'_{k_1}(\theta_{k_1}) + c_{k_1}] \cap [t_{d,\alpha'} - \eta'_k(\theta_k) < T_k \right. \\
&\quad \left. < t_{d,\alpha'} - \eta'_k(\theta_k) + c_k ] \cap \left[ \cap_{j=1}^{m, \neq k, k_1} (T_j > t_{d,\alpha'} - \eta'_j(\theta_j)) \right] \right\}, \quad (30)
\end{aligned}
$$

which clearly is a monotone increasing function of $\theta_{k_1}$ conditional on $\theta^{(k_1)}$, regardless of $c_k$ for $k = 1, \ldots, m$. □

## A.2 Proof of Theorem 1

*Proof*: Recall that $H_{0k}^{NI}$ implies $H_{0k}^{S}$ and that the null hypothesis $H_0 : \left(\cap_{k=1}^{m} H_{0k}^{S}\right) \cup \left(\cup_{k=1}^{m} H_{0k}^{NI}\right)$ can be decomposed into the following all possible individual components:

$$H_{0i} = \begin{cases} \cap_{k=1}^{m} H_{0k}^{S} = H_0^{S}, \\ H_0^{S(-k)} H_{0k}^{NI}, & k = 1, \ldots, m, \\ H_0^{S(-k_1, -k_2)} H_{0k_1}^{NI} H_{0k_2}^{NI}, & 1 \leq k_1 < k_2 \leq m, \\ \vdots \\ H_{0k}^{S} H_0^{NI(-k)}, & k = 1, \ldots, m \\ H_{0k}^{NI}, & k = 1, \ldots, m, \\ H_{0k_1}^{NI} H_{0k_2}^{NI}, & 1 \leq k_1 < k_2 \leq m, \\ \vdots \\ H_{01}^{NI} \ldots H_{0m}^{NI} = H_0^{NI} \end{cases} \quad (31)$$

where $H_0^{S(-k)} = \cap_{i=1}^{m, \neq k} H_{0i}^{S}$, $H_0^{S(-k_1, -k_2)} = \cap_{i=1}^{m, \neq k_1, k_2} H_{0i}^{S}$, and $H_0^{NI(-k)} = \cap_{i=1}^{m, \neq k} H_{0i}^{NI}$. Let $\gamma(H_{0i})$ be the upper bound of the Type I error probability under, i.e., the right-hand side of (7), under one of the null hypothesis configurations listed in (31), denoted by $H_{0i}$. As $\gamma(H_{0i})$ is monotone with respect to $\theta_k$ while conditioning on $\theta^{(-k)}$, one has the following

$$\begin{aligned} \gamma(\cap_{k=1}^{m} H_{0k}^{S}) &\geq \gamma'(H_0^{S(-k)} H_{0k}^{NI}) \\ &\geq \gamma(H_0^{S(-k_1, -k_2)} H_{0k_1}^{NI} H_{0k_2}^{NI}) \\ &\geq \ldots \\ &\geq \gamma(H_{0k}^{S} H_0^{NI(-k)}) \end{aligned} \quad (32)$$

and

$$\begin{aligned} \gamma(H_{0k}^{NI}) &\geq \gamma(H_{0k_1}^{NI} H_{0k_2}^{NI}) \\ &\geq \ldots \\ &\geq \gamma(H_{01}^{NI} \ldots H_{0m}^{NI}). \end{aligned} \quad (33)$$

That is, the maximum of the upper bound of the Type I error rate is either $\gamma(\cap_{k=1}^{m} H_{0k}^{S})$ or $\max_{1 \leq k \leq m} \gamma(H_{0k}^{NI})$. Note that under $H_{0}^{S} = \cap_{k=1}^{m} H_{0k}^{S}$, one has

$$\gamma(\cap_{k=1}^{m} H_{0k}^{S}) \leq \sum_{k=1}^{m} P_0\{(T_k > t_{d,a'}) \cap (\cap_{i=1, i \neq k}^{m}(T_i > t_{d,a'} - c_i))\} = \gamma_1(\alpha') \qquad (34)$$

and under $H_{0k}^{NI}$ one has

$$\gamma(H_{0k}^{NI}) \leq \max_{1 \leq k \leq m} P_0\{T_k > t_{d,a'} + c_k\} + (m-1)P_0\{T_k > t_{d,a'}\} = \gamma_2(\alpha'). \qquad (35)$$

Whether $\gamma_1(\alpha') \geq \gamma_2(\alpha')$ or $\gamma_1(\alpha') \leq \gamma_2(\alpha')$ depends on the data structure (i.e., the number of endpoints $m$, partial correlation coefficient $\rho_{k_1,k_2}$ and degree of freedom $d$) as well as the pre-defined effect margin $c_k$. If one controls $\max(\gamma_1, \gamma_2)$, then the Type I error rate is controlled. This completes the proof.